\documentclass[floatfix,twocolumn,showpacs,amssymb,amsmath,prd]{revtex4}
\usepackage{graphicx}
\usepackage{dcolumn}
\usepackage{bm}

\begin{document}
\title{Anomalous parity asymmetry of WMAP 7 year power spectrum data at low multipoles: is it cosmological or systematics?}
\author{Jaiseung Kim}
\email{jkim@nbi.dk}
\affiliation{Niels Bohr Institute \& Discovery Center, Blegdamsvej 17, DK-2100 Copenhagen, Denmark}
\author{Pavel Naselsky}
\affiliation{Niels Bohr Institute \& Discovery Center, Blegdamsvej 17, DK-2100 Copenhagen, Denmark}

\date{\today}

\begin{abstract}
It is natural to assume a parity neutral Universe, and accordingly no particular parity preference in CMB sky.
However, our investigation based on the WMAP 7 year power spectrum shows there exist large-scale odd-parity preference with high statistical significance.
We also find the odd-parity preference in WMAP7 data is slightly higher than earlier releases.
We have investigated possible origins, and ruled out various non-cosmological origins.
We also find primordial origin requires $|\mathrm{Re} [\Phi(\mathbf k)]|\ll|\mathrm{Im} [\Phi(\mathbf k)]|$ for $k\lesssim 22/\eta_0$,  where $\eta_0$ is the present conformal time.
In other words, it requires translational invariance in primordial Universe to be violated on the scales larger than $4\,\mathrm{Gpc}$. The Planck surveyor, which possesses wide frequency coverage and systematics distinct from the WMAP, may allow us to resolve the mystery of the anomalous odd-parity preference.
Furthermore, polarization maps of large sky coverage will reduce degeneracy in cosmological origins.
\end{abstract}

\pacs{95.85.Sz, 98.70.Vc, 98.80.Cq, 98.80.Es, 98.80.-k}
\maketitle 

\section{Introduction}
For the past years, there have been great successes in measurement of CMB anisotropy by ground and satellite observations  \citep{WMAP7:powerspectra,WMAP7:basic_result,WMAP5:basic_result,WMAP5:powerspectra,WMAP5:parameter,ACBAR,ACBAR2008,QUaD1,QUaD2,QUaD:instrument,Planck_bluebook}.
CMB anisotropy, which are associated with the inhomogeneity of the last scattering surface, provides the deepest survey so far and allows us to 
constrain cosmological models significantly. 
Since the initial release of WMAP data, the WMAP CMB sky data have undergone scrutiny, and various anomalies have been found and reported \citep{cold_spot1,cold_spot2,cold_spot_wmap3,cold_spot_origin,Tegmark:Alignment,Multipole_Vector1,Multipole_Vector2,Multipole_Vector3,Multipole_Vector4,Axis_Evil,Axis_Evil2,Axis_Evil3,Universe_odd,Park_Genus,Chiang_NG,Hemispherical_asymmetry,power_asymmetry_subdegree,power_asymmetry_wmap5,alfven,fnl_power,odd,lowl_anomalies}. 
In particular, CMB anisotropy at low multipoles are associated with scales far beyond any existing astrophysical survey, and therefore 
CMB anomalies at low multipoles may hint new physical laws at unexplored large scales. 

CMB sky map may be considered as the sum of even and odd parity functions. Previously, Land and et al. have noted the odd point-parity preference of WMAP CMB data, but found its statistical significance is not high enough \citep{Universe_odd}.
In our previous work, we have applied a slight different estimator to WMAP 5 year power spectrum data, and found 
significant odd point-parity preference at low multipoles \citep{odd}.
In this paper, we investigate the recently released WMAP 7 year data up to higher multipoles, and discuss origins of the observed odd-parity preference.

This paper is organized as follows.
In Section \ref{asymmetry}, we discuss the anomalous odd-parity preference of the WMAP data.
In Section \ref{cosmomc}, we implement cosmological fitting by excluding even or odd low multipole data.
In Section \ref{systematics} and \ref{cosmic}, we investigate non-cosmological and cosmological origin.
In Section \ref{discussion}, we summarize our investigation and discuss prospect.
In Appendix \ref{CMB}, we briefly review statistical properties of Gaussian CMB anisotropy. 

\section{Parity asymmetry of the WMAP data}
\label{asymmetry}
\begin{figure}[htb!]
\centering\includegraphics[scale=.5]{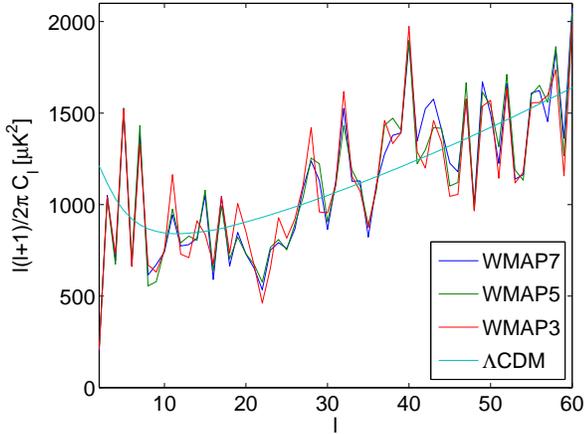}
\caption{CMB power spectrum: WMAP 7 year data (blue), WMAP 5 year data (green) and WMAP 3 year data (red), $\Lambda$CDM model (cyan)}
\label{Cl_cut}
\end{figure}

CMB anisotropy sky map may be considered as the sum of even and odd parity functions:
\begin{eqnarray} 
T(\hat{\mathbf n})=T^+(\hat{\mathbf n})+T^-(\hat{\mathbf n}),  
\end{eqnarray}
where
\begin{eqnarray} 
T^+(\hat{\mathbf n})&=&\frac{T(\hat{\mathbf n})+T(-\hat{\mathbf n})}{2},\\
T^-(\hat{\mathbf n})&=&\frac{T(\hat{\mathbf n})-T(-\hat{\mathbf n})}{2}.
\end{eqnarray}
Taking into account the parity property of spherical harmonics $Y_{lm}(\hat{\mathbf n})=(-1)^l\,Y_{lm}(-\hat{\mathbf n})$ \citep{Arfken},
we may easily show
\begin{eqnarray} 
T^+(\hat{\mathbf n})&=&\sum_{l=2n,m} a_{lm}\,Y_{lm}(\hat{\mathbf n}), \label{T_even}\\
T^-(\hat{\mathbf n})&=&\sum_{l=2n-1,m} a_{lm}\,Y_{lm}(\hat{\mathbf n}), \label{T_odd}
\end{eqnarray}
where $n$ is an integer.
Therefore, significant power asymmetry between even and odd multipoles indicates a preference for a particular parity. 
Hereafter, we will denote a preference for particular parity by `parity asymmetry'. 
In Fig. \ref{Cl_cut}, we show the WMAP 7 year, 5 year, 3 year data and the WMAP concordance model  \citep{WMAP7:powerspectra,WMAP5:powerspectra,WMAP3:temperature,WMAP5:Cosmology,WMAP7:Cosmology}. 
We may see from Fig. \ref{Cl_cut} that the power spectrum of WMAP data at even multipoles tend to be lower than those at neighboring odd multipoles. 
\begin{figure}[htb]
\centering\includegraphics[scale=.5]{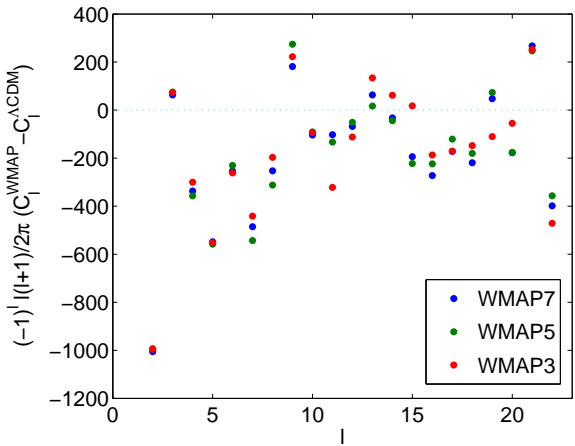}
\caption{$(-1)^l\times$ difference between WMAP power spectrum data and $\Lambda$CDM model}
\label{delta}
\end{figure}
In Fig. \ref{delta}, we show $(-1)^l l(l+1)/2\pi\:(C^{\mathrm{WMAP}}_l-C^{\Lambda\mathrm{CDM}}_l)$ for low multipoles.
Since we expect random scattering of data points around a theoretical model, we expect the distribution of dots in Fig. \ref{delta} to be symmetric around zero.
However, there are only 5 points of positive values among 22 points in the case of WMAP7 or WMAP5 data.
Therefore, we may see that there is the tendency of power deficit (excess) at even (odd) multipoles, compared with the $\Lambda$CDM model.  
Taking into account $l(l+1) C_l\sim \mathrm{const}$, we may consider the following quantities:
\begin{eqnarray} 
P^{+} &=& \sum^{l_{\mathrm{max}}}_{l=2} 2^{-1}\left(1+(-1)^{l}\right)\, l(l+1)/2\pi \: C_l,\\
P^{-} &=& \sum^{l_{\mathrm{max}}}_{l=2} 2^{-1}\left(1-(-1)^{l}\right)\, l(l+1)/2\pi \: C_l,
\end{eqnarray}
where $P^{+}$ and $P^{-}$ are the sum of $l(l+1)/2\pi \: C_l$ for even and odd multipoles respectively. 
Therefore, the ratio $P^{+}/P^{-}$ is associated with the degree of the parity asymmetry, where the lower value of $P^+/P^-$ indicates odd-parity preference, and vice versa. It is worth to note that our estimator of the parity asymmetry does not possess explicit dependence on the underlying theoretical model in the sense that the term $C^{\mathrm{WMAP}}_l-C^{\Lambda\mathrm{CDM}}_l$ is absent in our estimator.

\begin{figure}[htb]
\centering\includegraphics[scale=.5]{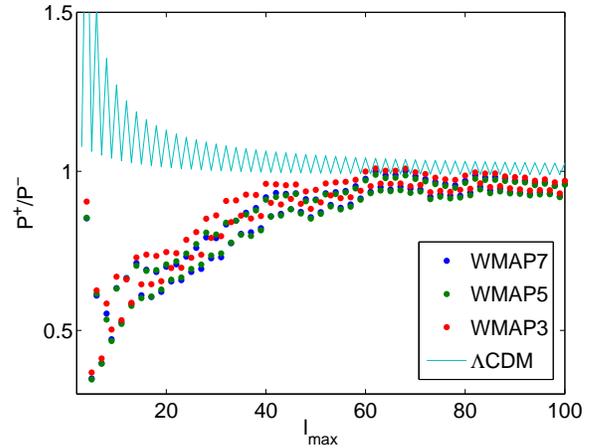}
\caption{$P^+/P^-$ of WMAP data and $\Lambda$CDM}
\label{P_ratio}
\end{figure}
In Fig. \ref{P_ratio}, we show the $P^+/P^-$ of WMAP data, and a $\Lambda$CDM model for various $l_{\mathrm{max}}$.
As shown in Fig. \ref{P_ratio}, $P^+/P^-$ of WMAP data are far below theoretical values. Though the discrepancy is largest at lowest $l_{\mathrm{max}}$, its statistical significance is not necessarily high for low $l$, due to large statistical fluctuation.
In order to make rigorous assessment on its statistical significance at low $l$, we are going to compare $P^+/P^-$ of WMAP data with simulation.
We have produced $10^4$ simulated CMB maps HEALPix Nside=8 and Nside=512 respectively, via map synthesis with $a_{lm}$ randomly drawn from Gaussian $\Lambda$CDM model.
We have degraded the WMAP processing mask (Nside=16) to Nside=8, and set pixels to zero, if any of their daughter pixels is zero. 
After applying the mask, we have estimated power spectrum $2\le l\le 23$ from simulated cut-sky maps (Nside=8) by a pixel-based maximum likelihood method \citep{WMAP7:powerspectra,Bond:likelihood,hybrid_estimation}. 
At the same time, we have applied the WMAP team's KQ85 mask to the simulated maps (Nside=512), and estimated power spectrum $2\le l\le 1024$  by pseudo $C_l$ method \citep{pseudo_Cl,MASTER}. In the simulation, we have neglected instrument noise, since the signal-to-noise ratio of the WMAP data is quite high at multipoles of interest (i.e. $l\le 100$) \citep{WMAP7:powerspectra,WMAP7:basic_result}. Using the low $l$ estimation by pixel-maximum likelihood method and high $l$ estimation by pseudo $C_l$ method, we have computed $P^+/P^-$ respectively for various multipole ranges $2\le l \le l_{\mathrm{max}}$, and compared $P^+/P^-$ of the WMAP data with simulation.
\begin{figure}[htb]
\centering\includegraphics[scale=.5]{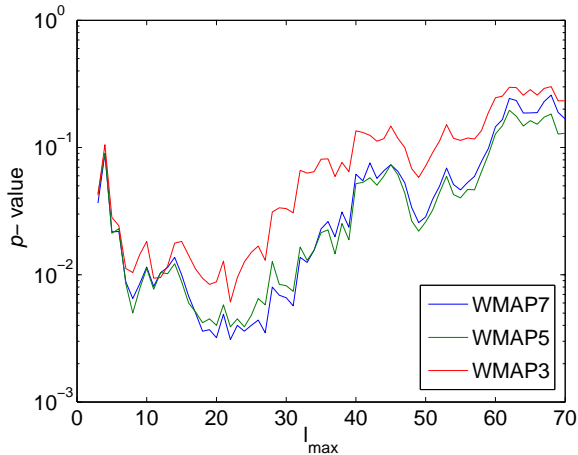}
\caption{Probability of getting $P^+/P^-$ as low as WMAP data for multipole range $2\le l\le{l_\mathrm{max}}$.}
\label{P}
\end{figure}
In Fig. \ref{P}, we show $p$-value of WMAP7, WMAP5 and WMAP3 respectively for various $l_{\mathrm{max}}$, where $p$-value denotes fractions of simulations as low as $P^+/P^-$ of the WMAP data. As shown in Fig. \ref{P}, the parity asymmetry of WMAP7 data at multipoles ($2\le l\le 22$) is most anomalous, where $p$-value is $0.0031$.
As shown in Fig. \ref{P}, the statistical significance of the parity asymmetry (i.e. low $p$-value) is getting higher, when we include higher multipoles up to 22.
Therefore, we may not attribute the odd parity preference simply to the low quadrupole power, and find it rather likely that the low quadrupole power is not an isolated anomaly, but shares an origin with the odd parity preference. 

\begin{figure}[htb]
\centering\includegraphics[scale=.5]{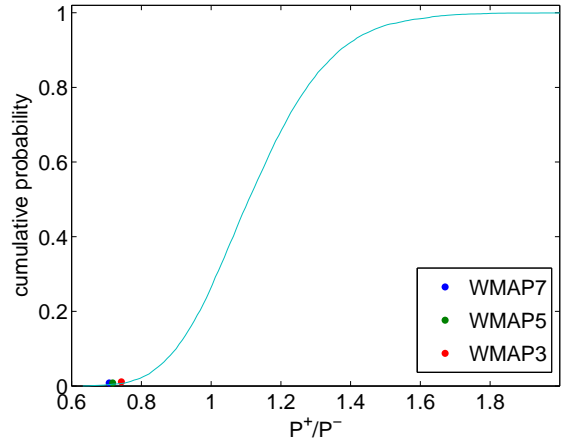}
\caption{Parity asymmetry at multipoles ($2\le l\le 22$): cumulative distribution of $P^+/P^-$ for $10^4$ simulated maps (cyan), $P^+/P^-$ of WMAP7 (blue), WMAP5 (green) and WMAP3 (red)}
\label{hist}
\end{figure}

\begin{table}[htb]
\centering
\caption{the parity asymmetry of WMAP data ($2\le l\le 22$)}
\begin{tabular}{ccc}
\hline\hline 
data &  $P^+/P^-$  &  $p$-value\\
\hline
WMAP7  &  0.7076  & 0.0031\\
WMAP5  &  0.7174 &  0.0039\\
WMAP3  &  0.7426 & 0.0061\\
\hline
\end{tabular}
\label{pvalue}
\end{table}
In Table \ref{pvalue}, we summarize $P^+/P^-$ and $p$-values of WMAP7, WMAP5 and WMAP3 for $l_{\mathrm{max}}=22$. As shown in Fig. \ref{P} and Table \ref{pvalue}, the odd-parity preference of WMAP7 is most anomalous, while WMAP7 data are believed to have more accurate calibration and less foreground contamination than earlier releases.  \citep{WMAP7:powerspectra,WMAP7:basic_result,WMAP5:basic_result,WMAP5:powerspectra,WMAP5:beam}. 
In Fig. \ref{hist}, we show cumulative distribution of $P^+/P^-$ for $10^4$ simulated maps. The values corresponding to $P^+/P^-$ of WMAP data are marked as dots.
We have also compared $P^+/P^-$ of the WMAP7 with whole-sky simulation (i.e. no mask), and obtained $p$-value $0.002$ for $l_{\mathrm{max}}=22$.
The lower $p$-value from whole-sky simulation is attributed to the fact that statistical fluctuation in whole-sky $C_l$ estimation is smaller than that of cut-sky estimation. 

In the absence of strong theoretical grounds for the parity asymmetry ($2\le l\le 22$),
we have to take into account our posteriori choice on $l_{\mathrm{max}}$, which might have enhanced the statistical significance.
In order to do that, we have produced whole-sky Monte-Carlo simulations, and retained only simulations, whose $p$-value is lowest at $l_{\mathrm{max}}=22$.
The $p$-values have been estimated by comparing them with simulations, which are produced separately.
Once we have retained $10^4$ simulations, we have compared them with WMAP7 data, and found that only fraction $0.0197$ of retained simulations have $P^+/P^-$ as low as WMAP7 data. 
The statistical significance of the parity asymmetry ($2\le l\le 22$) is reduced substantially by accounting for the posteriori choice on $l_{\mathrm{max}}$.
However, it is still significant.

We have also investigated the parity asymmetry with respect to mirror reflection (i.e. $l+m$=even or odd) respectively in Galactic coordinate and Ecliptic coordinate. However, we find the statistical significance is not as high as the point-parity asymmetry (i.e. $l$=even or odd).

\section{the power contrast and the $\Lambda$CDM model fitting}
\label{cosmomc}
The parity asymmetry discussed in the previous section is explicitly associated with the angular power spectrum data, which are used extensively to fit cosmological models.
Noting the significant power contrast between even and odd low multipoles, we have investigated cosmological models respectively by excluding even or odd low multipole data ($2\le l \le 22$) from total dataset.
Hereafter, we denote CMB data of even(odd)  multipoles ($2\le l\le 22$) plus all high multipoles as `$D_2$'(`$D_3$') respectively.
For a cosmological model, we have considered $\Lambda$CDM + SZ effect + weak-lensing, where cosmological parameters are $\lambda \in \left\{\Omega_b,\Omega_{c},\tau,n_s, A_s, A_{sz}, H_0 \right\}$. For data constraints, we have used the WMAP 7 year power spectrum data \citep{WMAP7:powerspectra}.
In order to exclude even or odd low multipoles, we have made slight modifications to the likelihood code provided by the WMAP team, and run a \texttt{CosmoMC} with the modification. \citep{Gibbs_power,WMAP7:powerspectra,CosmoMC}. 
\begin{figure}[htb]
\centering
\includegraphics[scale=.54]{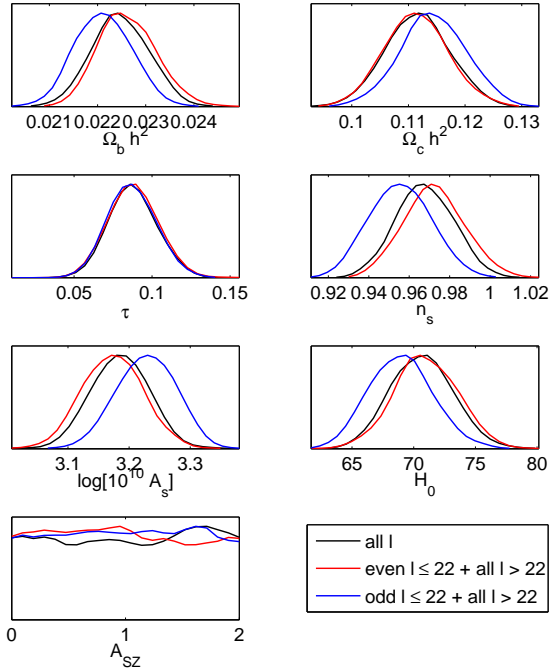}
\caption{Marginalized likelihood of cosmological parameters:  results are obtained respectively with/without even or odd multipole data ($2\le l \le 22$).}
\label{like}
\end{figure}
In Fig. \ref{like}, we show the marginalized likelihoods of parameters.
As shown in Fig. \ref{like}, the parameter likelihood imposed by $D_3$ seem to differ from those of others.
In Table \ref{parameter}, we show the bestfit parameters and 1 $\sigma$ confidence intervals, where $\lambda$, $\lambda_2$ and $\lambda_3$ denote the bestfit values of the full data, $D_2$ and $D_3$ respectively.
\begin{table}[htb]
\centering
\caption{cosmological parameters ($\Lambda$CDM + sz + lens)}
\begin{tabular}{cccc}
\hline\hline 
 & $\lambda$  &$\lambda_2$   & $\lambda_3$ \\
\hline
$\Omega_{b}\,h^2$  & $0.0226\pm 0.0006$ &$0.0228\pm0.0006$ & $0.0221\pm0.0006$ \\ 
$\Omega_{c}\,h^2$  & $0.112\pm0.006$ &$0.11\pm0.006$ & $0.116\pm0.006$ \\ 
$\tau$  & $0.0837\pm 0.0147$ &$0.0879\pm0.015$ & $0.087\pm0.0147$ \\ 
$n_s$ & $0.964\pm 0.014$ &$0.974\pm0.015$ & $0.95\pm0.015$ \\ 
$\log[10^{10} A_s]$  & $3.185\pm0.047$ &$ 3.165\pm0.049$ & $3.246\pm0.048$ \\ 
$H_0$  & $70.53\pm2.48$ &$71.43\pm2.51$ & $68.07\pm2.53$ \\
$A_{\mathrm{sz}}$  & $1.891^{+0.109}_{-1.891}$ &$1.469^{+0.541}_{-1.469}$ & $1.558^{+0.442}_{-1.558}$ \\ 
\hline
\end{tabular}
\label{parameter}
\end{table}
As mentioned above, there exist some level of tension between $D_3$ and the full data.
However, the Bayes Factor $\mathcal{L}(\lambda|D_3)/\mathcal{L}(\lambda_3|D_3)= \exp(-3733.124+3732.791)=0.71$ shows that
it is `not worth than a bare mention', according to Jeffreys' scale \citep{Jeffreys_scale}.

\section{Non-cosmological origins}
\label{systematics}
In this section, we are going to investigate non-cosmological origins such as asymmetric beams, instrument noise, foreground and cut-sky effect.

\subsection{asymmetric beam}
\begin{figure}[htb]
\centering
\includegraphics[scale=.25]{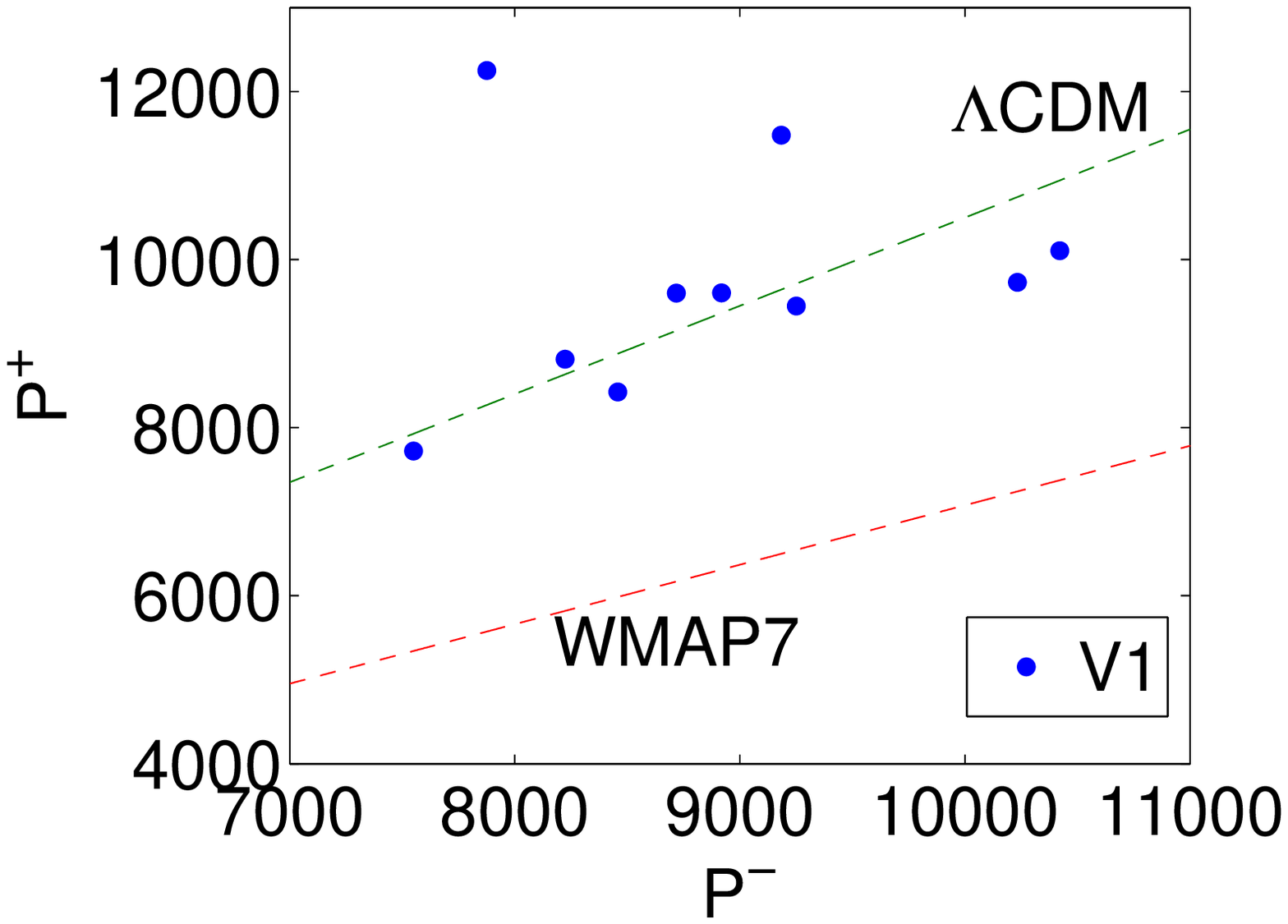}
\includegraphics[scale=.25]{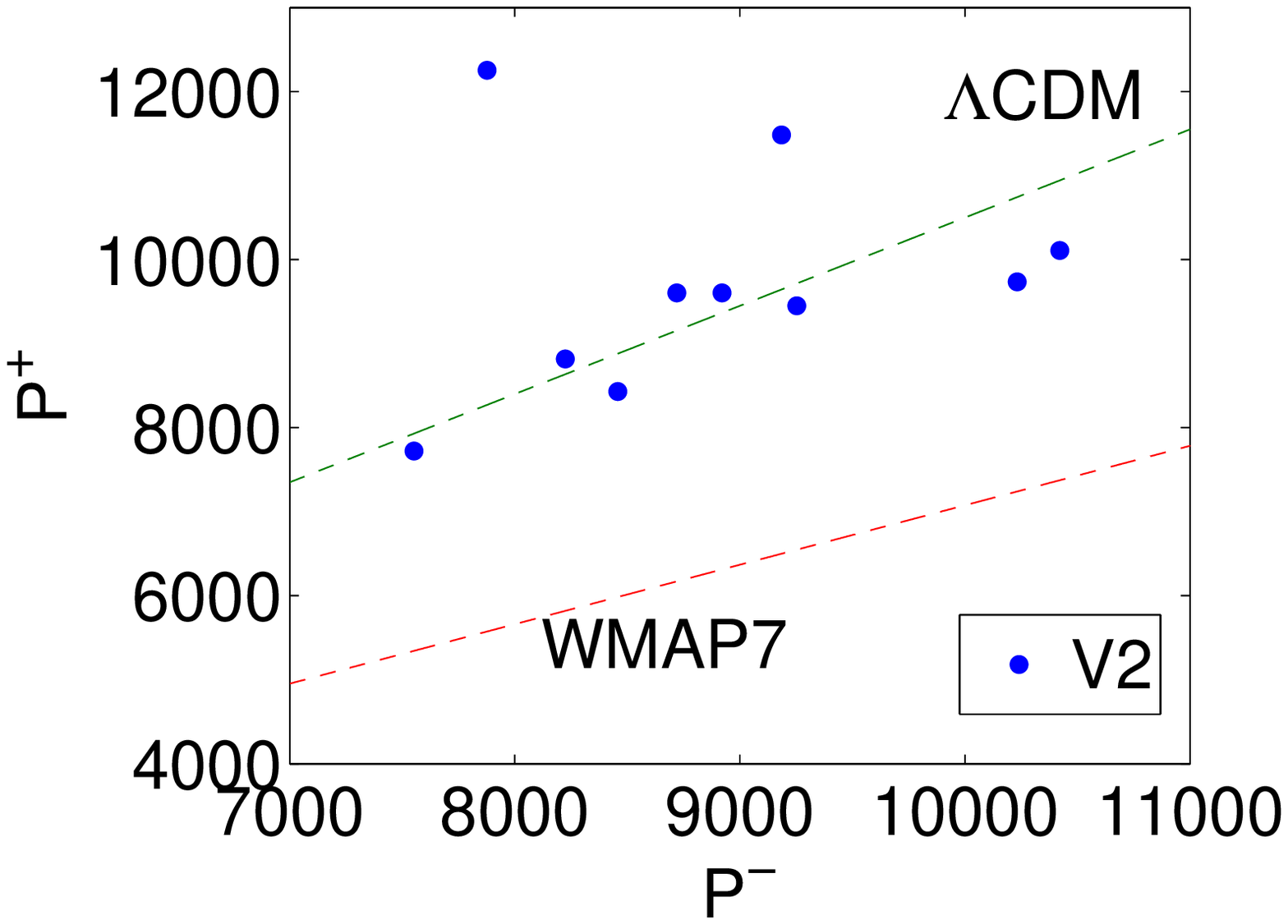}
\includegraphics[scale=.25]{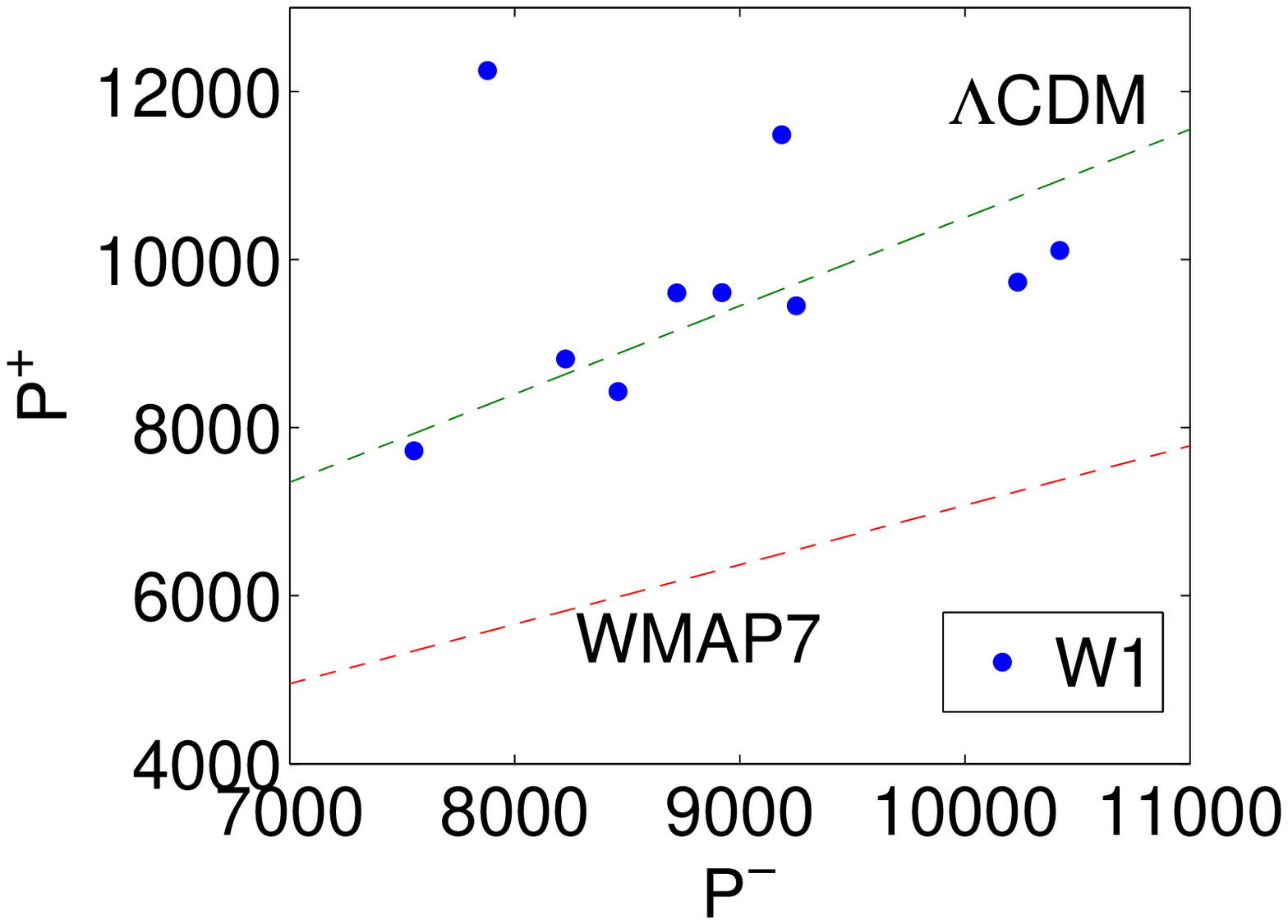}
\includegraphics[scale=.25]{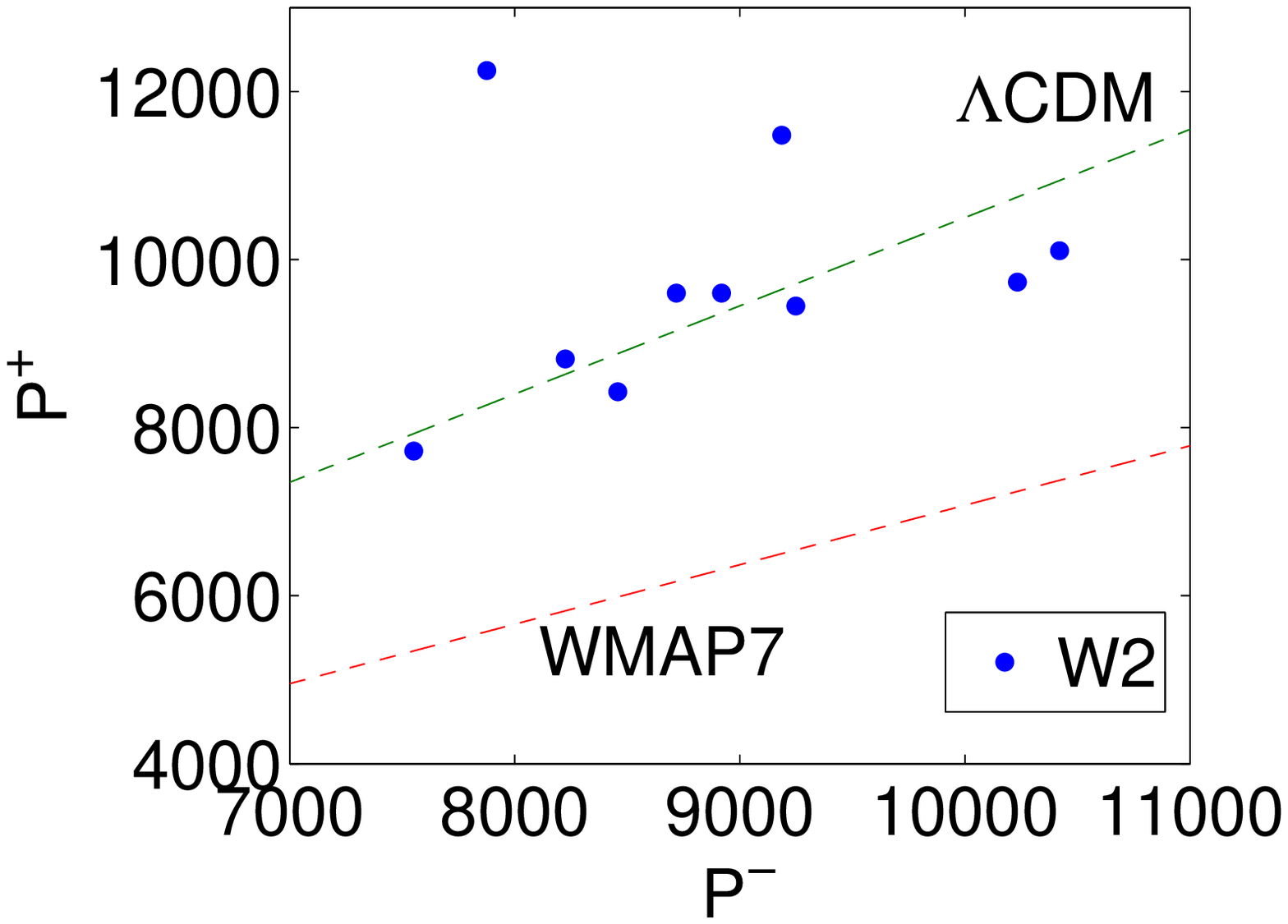}
\includegraphics[scale=.25]{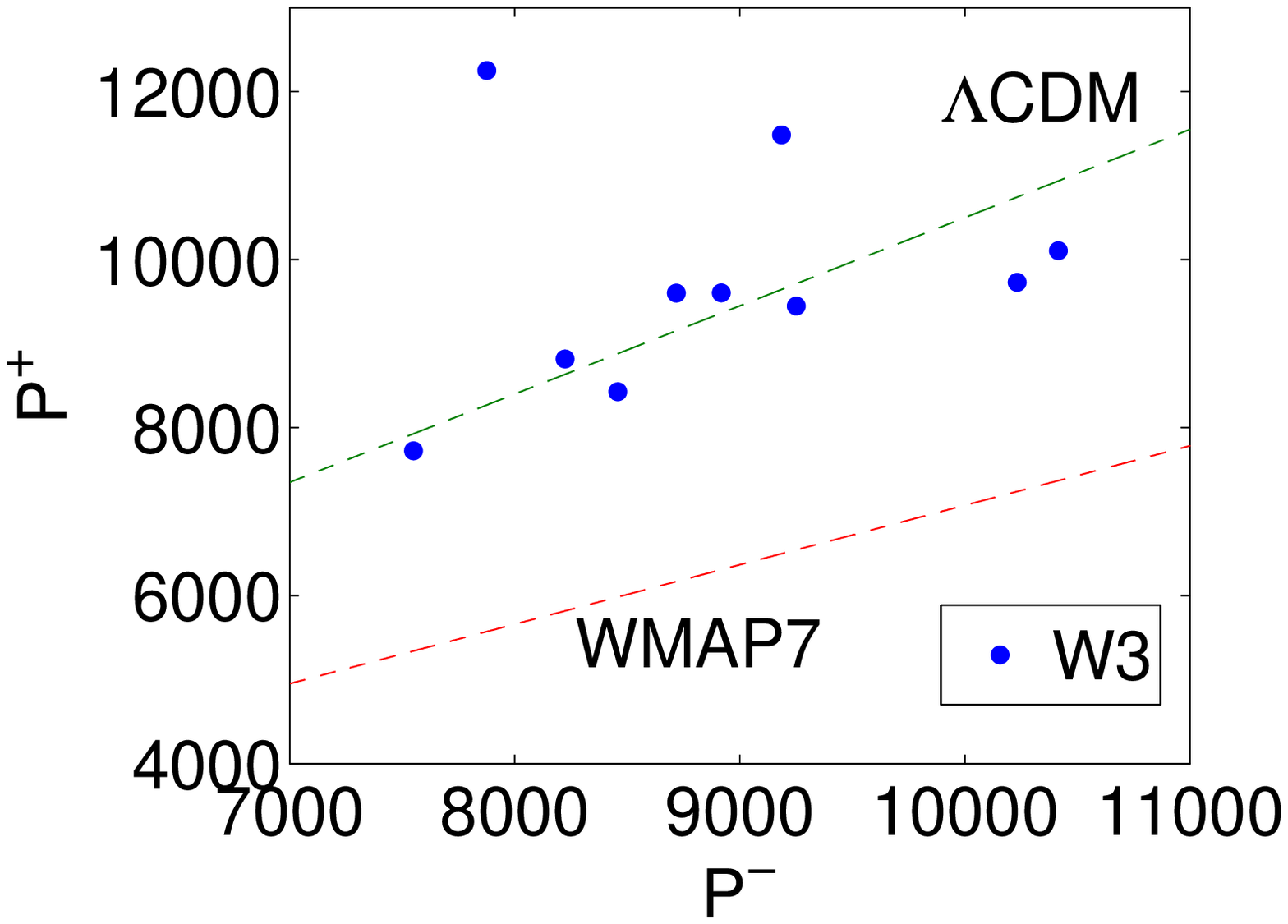}
\includegraphics[scale=.25]{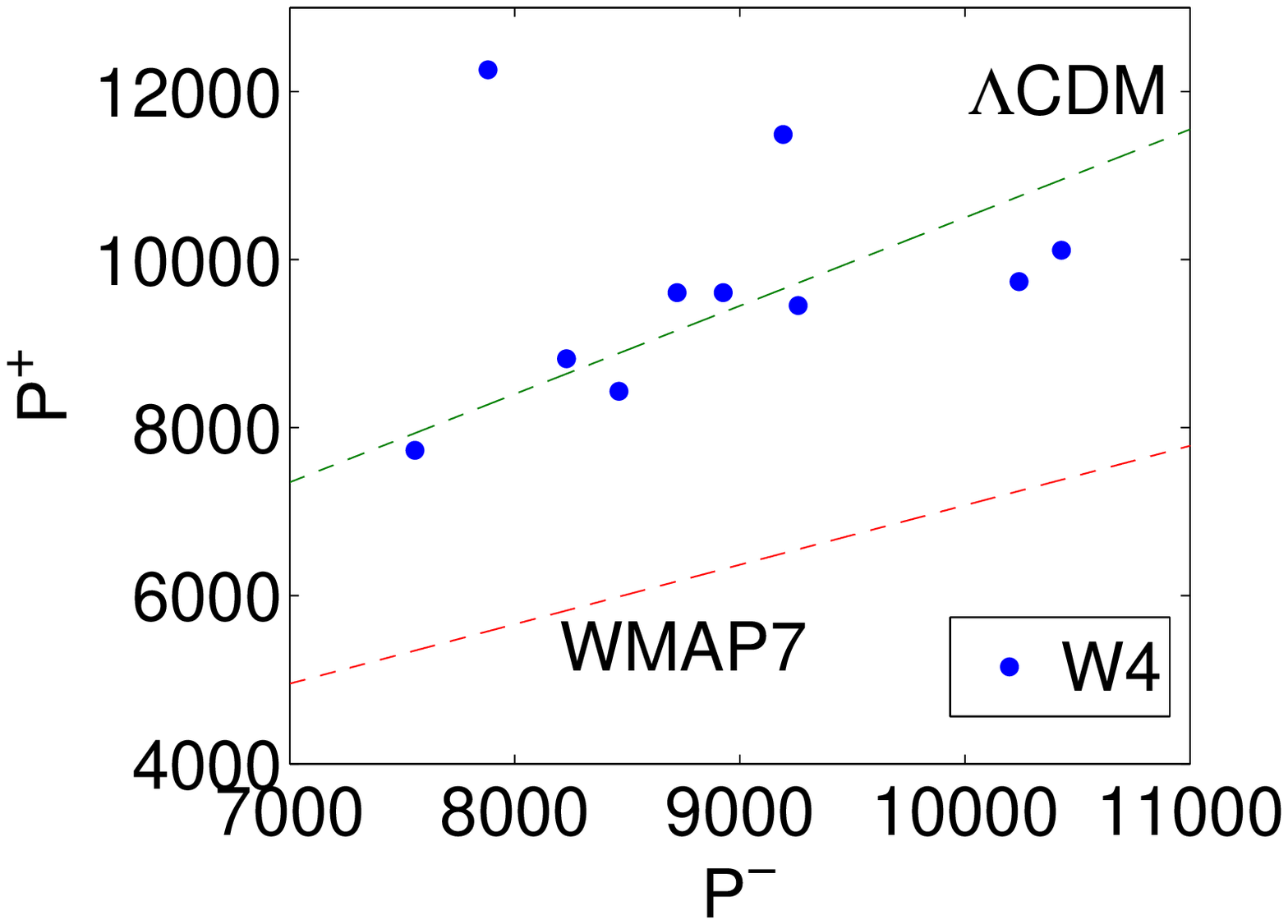}
\caption{the parity asymmetry in the presence of beam asymmetry: 
The dots denotes ($P^+$,$P^-$) of CMB maps simulated with asymmetric beams. 
The dashed lines are plotted with slopes corresponding to the $P^+/P^-$ of $\Lambda$CDM model (red) and WMAP7 data (green) respectively.
The alphanumeric values at the lower right corner denote the frequency band and D/A channel.}
\label{beam}
\end{figure} 
The shape of the WMAP beams are slightly  asymmetric \citep{WMAP5:beam,WMAP3:beam,asymmetric_beam}, while
the WMAP team have assumed symmetric beams in the power spectrum estimation \citep{WMAP7:powerspectra,WMAP5:powerspectra,WMAP5:beam,WMAP3:beam}.
We have investigated the association of beam asymmetry with the anomaly, by relying on 
simulated maps provided by \citep{asymmetric_beam}.
The authors have produced 10 simulated maps for each frequency and Differencing Assembly (D/A) channels, where the detailed shape of the WMAP beams and the WMAP scanning strategy
are taken into account \citep{asymmetric_beam}.
From simulated maps, we have estimated $P^+$ and $P^-$, where we have compensated for beam smoothing purposely by the WMAP team's beam transfer function (i.e. symmetric beams).
In Fig. \ref{beam}, we show $P^+$ and $P^-$ values of the simulated maps, and the dashed lines of a slope corresponding to $P^+/P^-$ of $\Lambda$CDM and WMAP7 data respectively.
As shown in Fig. \ref{beam}, we do not observe the odd-parity preference of WMAP data in simulated maps.
Therefore, we find it hard to attribute the odd-parity preference to asymmetric beams.

\subsection{noise}
\begin{figure}[htb]
\centering
\includegraphics[scale=.5]{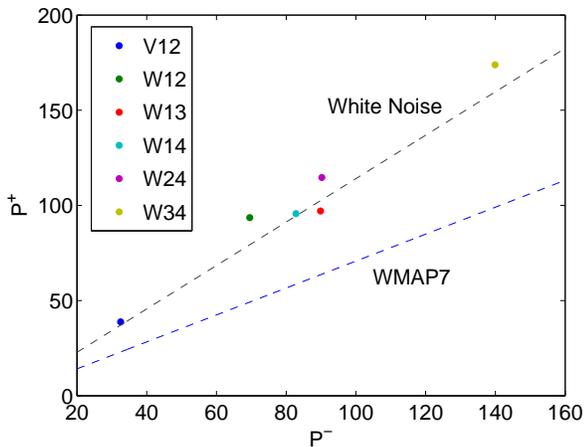}
\caption{the parity asymmetry of the WMAP noise: 
The dots denotes ($P^+$,$P^-$) of noise maps, and alphanumeric values in the legend denote the frequency band and the pair of D/A channels used. Two dashed lines are plotted with the slope corresponding to $P^+/P^-$ of white noise and WMAP7 data respectively.}
\label{noise}
\end{figure}

There exist instrument noise in the WMAP data.
Especially, 1/f noise, when coupled with WMAP scanning pattern, may result in less accurate measurement at certain low multipoles  \citep{WMAP3:temperature,Detection_Light,WMAP1:processing}. 
In order to investigate the association of noise with the anomaly, we have produced noise maps of WMAP7 data by subtracting one Differencing Assembly (D/A) map from another D/A data of the same frequency channel. 
In Fig. \ref{noise}, we show $P^+$ and $P^-$ values of the noise maps.
As shown in Fig. \ref{noise}, the noise maps do not show odd-parity preference, but their $P^+/P^-$ ratios are consistent with that of white noise (i.e. $C_l=\mathrm{const.}$).
Besides that, the signal-to-noise ratio of WMAP temperature data is quite high at low multipoles (e.g S/N$\sim$ 100 for $l=30$)  \citep{WMAP3:temperature,WMAP5:beam,WMAP1:processing}. 
Therefore, we find that instrument noise, including 1/f noise, is unlikely to be the cause of the odd-parity preference.

\subsection{foreground}
\begin{figure}[htb]
\centering
\includegraphics[scale=.5]{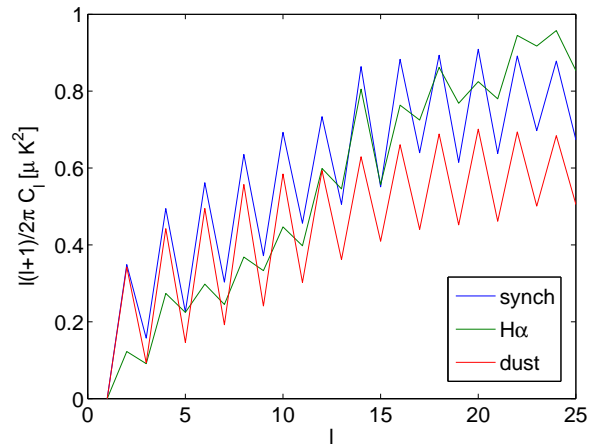}
\caption{the power spectra of the templates (synchrotron, H$\alpha$, dust): plotted with arbitrary normalization.}
\label{template}
\end{figure}

There are contamination from galactic and extragalactic foregrounds.
In order to reduce foreground contamination, the WMAP team have subtracted diffuse foregrounds by template-fitting, and masked the regions that cannot be cleaned reliably.
For foreground templates (dust, free-free emission and synchrotron), the WMAP team used dust emission ``Model 8", H$\alpha$ map, and the difference between K and Ka band maps  \citep{WMAP5:basic_result,WMAP7:fg,Dust_Extrapolation,Finkbeiner_H_alpha,WMAP1:fg}. 
In Fig. \ref{template}, we show the power spectrum of templates. 
As shown in Fig. \ref{template}, templates show strong even parity preference, which is opposite to that of the WMAP power spectrum data.
Therefore, one might attribute the odd-parity preference of WMAP data to over-subtraction by templates.
However, we find the error associated with template-fitting is unlikely to be the cause of the odd-parity preference.
Consider spherical harmonic coefficients of a foreground-reduced map:
\begin{eqnarray} 
a^{\mathrm{obs}}_{lm}=a^{\mathrm{cmb}}_{lm}+a^{\mathrm{fg}}_{lm}-b\,a^{\mathrm{tpl}}_{lm},
\end{eqnarray}
where $a^{\mathrm{obs}}_{lm}$, $a^{\mathrm{fg}}_{lm}$ and $b\,a^{\mathrm{tpl}}_{lm}$ correspond to a foreground-cleaned map, a foreground and
a template with a fitting coefficient $b$.
For simplicity, we consider only a single foreground component, but the conclusion is equally valid for multi-component foregrounds.
Since there is no correlation between foregrounds and CMB, the observed power spectrum is given by:
\begin{eqnarray} 
C^{\mathrm{obs}}_l\approx C^{\mathrm{cmb}}_l+ \langle \left|a^{\mathrm{fg}}_{lm}  -b\, a^{\mathrm{tpl}}_{lm} \right|^2\rangle. \label{Cl_obs}
\end{eqnarray}
As shown Eq. \ref{Cl_obs},  the parity preference should follow that of templates because of the second term, provided templates are good tracers of foregrounds (i.e. $a^{\mathrm{fg}}_{lm}/a^{\mathrm{tpl}}_{lm}\approx \mathrm {const}$). Nevertheless, Eq. \ref{Cl_obs} may make a bad approximation for lowest multipoles, because the cross term $\sum_m \mathrm{Re}[a^{\mathrm{cmb}}_{lm} (a^{\mathrm{fg}}_{lm}  -b\, a^{\mathrm{tpl}}_{lm})^*]$ may not be negligible. 
Besides that, our argument and the template-fitting method itself fail, if templates are not good tracers of foregrounds. 
In order to investigate these issues, we have resorted to simulation in combination with WMAP data.
Noting the WMAP power spectrum is estimated from foreground-reduced V and W band maps, we have produced simulated maps as follows:
\begin{eqnarray} 
T(\hat{\mathbf n})=T_{\mathrm{cmb}}(\hat{\mathbf n})+(V(\hat{\mathbf n})-W(\hat{\mathbf n}))/2,  \label{residual_fg2}
\end{eqnarray}
where $V(\hat{\mathbf n})$ and $W(\hat{\mathbf n})$ are foreground-reduced V and W band maps of WMAP data. Note that the second term on the right hand side contains only residual foregrounds at V and W band, because the difference of distinct frequency channels are mainly residual foregrounds at low $l$.
Just as cut-sky simulation described in Sec. \ref{asymmetry}, we have applied a foreground mask to the simulated maps, and estimated the power spectrum from cut-sky by a pixel-based maximum likelihood method.
\begin{figure}[htb]
\centering
\includegraphics[scale=.5]{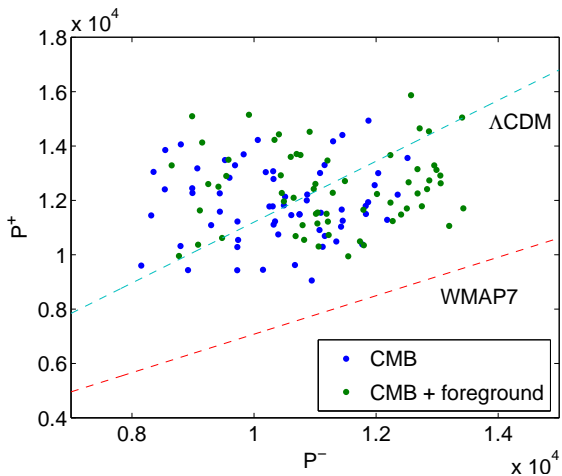}
\caption{the parity asymmetry in the presence of residual foregrounds (V$-$W): Dashed lines are plotted with slopes corresponding to $P^+/P^-$ of $P^+/P^-$ of $\Lambda$CDM (cyan), WMAP7 data (red).}
\label{residual2}
\end{figure} 
In Fig. \ref{residual2}, we show $P^+$ and $P^-$ values estimated from simulations. 
For comparison, we have included simulations without residual foregrounds, and dashed lines of a slope corresponding to $P^+/P^-$ of $\Lambda$CDM model and WMAP7 data. As shown in Fig. \ref{residual2}, the $P^+/P^-$ of simulations in the presence of residual foregrounds do not show anomalous odd-parity preference of WMAP data.
Considering Eq. \ref{Cl_obs} and simulations, we find it difficult to attribute the odd-parity preference to residual foreground.

There also exist contamination from unresolved extragalactic point sources \citep{WMAP3:temperature}. 
However, point sources follow Poisson distribution with little departure \citep{Tegmark:Foreground}, and therefore are unlikely to possess odd-parity preference.
Besides that, point sources at WMAP frequencies are subdominant on large angular scales (low $l$) \citep{WMAP3:temperature,WMAP7:fg,Tegmark:Foreground,WMAP5:fg}.

Though we have not find association of foregrounds with the anomaly, 
we do not rule out residual foreground completely, because of our limited knowledge on residual foregrounds.

\subsection{cut sky}
The WMAP team have masked the region that cannot be reliably cleaned by template fitting, and estimated CMB power spectrum from sky data outside the mask  \citep{WMAP7:powerspectra,WMAP5:powerspectra,WMAP3:temperature,WMAP7:fg}. 
Therefore, we have estimated the $p$-value presented in Sec. \ref{asymmetry}, by comparing WMAP data with cut-sky simulations.
\begin{figure}[htb]
\centering
\includegraphics[scale=.5]{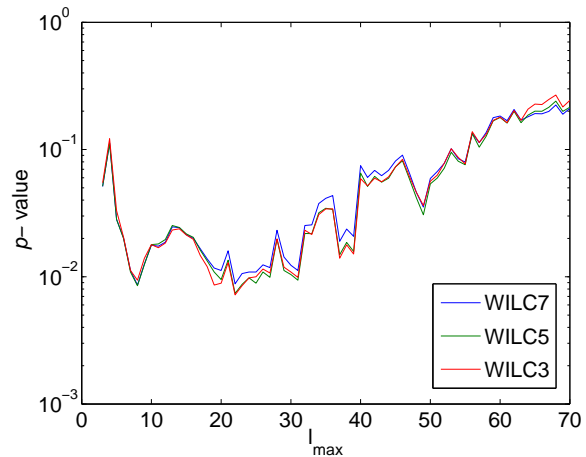}
\caption{Probability of getting $P^+/P^-$ as low as the ILC 7 year, 5 year, and 3 year map at multipole range $2\le l\le{l_\mathrm{max}}$}
\label{P_wilc}
\end{figure}
Nonetheless, we have investigated the WMAP team's Internal Linear Combination map (ILC) map in order to see if the odd-parity preference also exists in a whole-sky CMB map.
Note that the WMAP ILC map provides a reliable estimate of CMB signal over whole-sky on angular scales larger than $10^\circ$  \citep{WMAP3:temperature,WMAP7:fg,WMAP5:fg}.
We have compared $P^+/P^-$ of the ILC maps with whole-sky simulations.
In Fig. \ref{P_wilc}, we show $p$-values of the ILC maps respectively for various $l_{\mathrm{max}}$.
As shown in Fig. \ref{P_wilc}, the odd-parity preference of ILC maps is most anomalous for $l_{\mathrm{max}}=22$ as well.
In Table \ref{pvalue_wilc}, we summarize $P^+/P^-$ and $p$-values for $l_{\mathrm{max}}=22$. 

\begin{table}[htb]
\centering
\caption{the parity asymmetry of WMAP ILC maps ($2\le l\le 22$)}
\begin{tabular}{ccc}
\hline\hline 
data &  $P^+/P^-$  &  $p$-value\\
\hline
ILC7  &  0.7726  & 0.0088\\
ILC5  &  0.7673 &  0.0074\\
ILC3  &  0.7662 & 0.0072\\
\hline
\end{tabular}
\label{pvalue_wilc}
\end{table}

As shown in Fig. \ref{P_wilc} and Table \ref{pvalue_wilc}, we find anomalous odd-parity preference exits in whole-sky CMB maps as well.
Therefore, we find it difficult to attribute the anomaly to cut-sky effect.

\subsection{Other known sources of errors}
Besides contamination discussed in previous sections, there are other sources of contamination such as sidelobe pickup and so on.
In order to investigate these effects, we have resorted to simulation produced by the WMAP team.
According to the WMAP team, time-ordered data (TOD) have been simulated with realistic noise, thermal drifts in instrument gains and baselines,
smearing of the sky signal due to finite integration time, transmission imbalance, and far-sidelobe beam pickup.
Using the same data pipeline used for real data, the WMAP team have processed simulated TOD, and produced maps for each differencing assembly and each single year observation year.
\begin{figure}[htb]
\centering
\includegraphics[scale=.5]{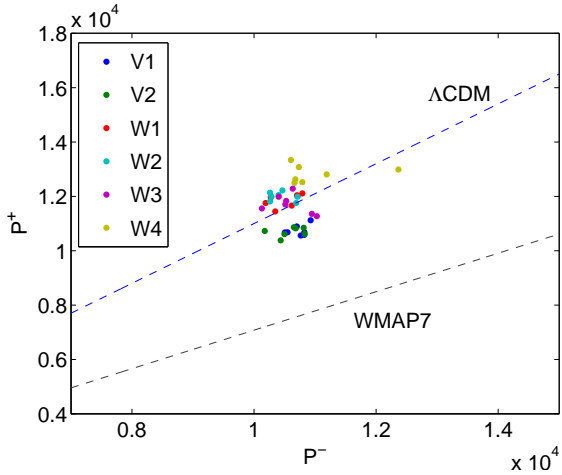}
\caption{ $P^+$ and $P^-$ of the WMAP team's simulation for V and W band data}
\label{sim}
\end{figure}
In Fig. \ref{sim}, we show the $P^+$ and $P^-$ of the simulated maps, where the power spectrum estimation is made from cut-sky by a pixel-based likelihood method. 
As shown in Fig. \ref{sim}, all points are well above $P^+/P^-$ of WMAP7, and agree with $\Lambda$CDM model.
Therefore, we do not find definite association of the parity asymmetry with known systematics effects.

\section{Cosmological origin}
\label{cosmic}
In this section, we are going to take the WMAP power spectrum at face values, and consider cosmological origins.
Topological models including multi-connected Universe and Bianchi $VII$ model have been proposed to explain the cold spot or low quadrupole power  \citep{low_quadrupole,spherical_tessellation,Land_Bianchi}.
However, the topological models do not produce the parity asymmetry, though some of them, indeed, predict low quadrupole power.
Trans-Planckian effects and some inflation models predict oscillatory features in primordial power spectrum  \citep{Inflation,Inflation_Planckian_problem,Inflation_Planckian_spectra,Inflation_Planckian_note,Inflation_Planckian_estimate,CMB_Planckian_signature,WMAP_oscillation,Inflation_Planckian,Inflation_initial,Planckian_Astrophysics,CMB_Planckian_observation,WMAP3:parameter}. 
However, oscillatory or sharp features in primordial power spectrum are smeared out in translation to the CMB power spectrum \citep{WMAP7:anomaly}.
Besides, reconstruction of primordial power spectrum and investigation on features show that primordial power spectrum is close to a featureless power-law spectrum \citep{WMAP7:powerspectra,WMAP5:Cosmology,WMAP7:Cosmology,WMAP3:parameter,power_recon,power_svd,power_features}.
Therefore, we find it difficult to attribute the anomaly to trans-Planckian effect or extended inflation models.
We will consider what the odd-parity preference imply on primordial perturbation $\Phi(\mathbf k)$, if primordial power spectrum is, indeed, featureless. Using Eq. \ref{alm}, we may show the decomposition coefficients of CMB anisotropy are given by: 
\begin{eqnarray*}
a_{lm}&=&\frac{(-\imath)^l}{2\pi^2} \int\limits^{\infty}_0 dk  \int\limits^{\pi}_0 d\theta_{\mathbf k} \sin\theta_{\mathbf k}  \int\limits^{2\pi}_0 d\phi_{\mathbf k}\,\Phi(\mathbf k)\,g_{l}(k)\,Y^*_{lm}(\hat{\mathbf k}),\nonumber\\
&=&\frac{(-\imath)^l}{2\pi^2} \int\limits^{\infty}_0 dk  \int\limits^{\pi}_0 d\theta_{\mathbf k} \sin\theta_{\mathbf k}\int\limits^{\pi}_0 d\phi_{\mathbf k}\,g_{l}(k)\times\\
&&\left(\Phi(\mathbf k)\,Y^*_{lm}(\hat {\mathbf k}) + \Phi(-\mathbf k)\,Y^*_{lm}(-\hat {\mathbf k}) \right),\nonumber\\
&=&\frac{(-\imath)^l}{2\pi^2} \int\limits^{\infty}_0 dk  \int\limits^{\pi}_0 d\theta_{\mathbf k} \sin\theta_{\mathbf k}\int\limits^{\pi}_0 d\phi_{\mathbf k}\,g_{l}(k) Y^*_{lm}(\hat {\mathbf k})\times\\
&&\left(\Phi(\mathbf k)+(-1)^l \Phi^*(\mathbf k)\right),
\end{eqnarray*}
where we used the reality condition $\Phi(-\mathbf k)= \Phi^*(\mathbf k)$ and $Y_{lm}(\hat{-\mathbf n})=(-1)^l\,Y_{lm}(\hat{\mathbf n})$.
Using Eq. \ref{alm2},  it is trivial to show, for the odd number multipoles $l=2n-1$,
\begin{eqnarray}
\lefteqn{a_{lm}=}\label{alm2}\\
&&-\frac{(-\imath)^{l-1}}{\pi^2} \int\limits^{\infty}_0 dk  \int\limits^{\pi}_0 d\theta_{\mathbf k} \sin\theta_{\mathbf k}\int\limits^{\pi}_0 d\phi_{\mathbf k}\,g_{l}(k) Y^*_{lm}(\hat {\mathbf k})\,\mathrm{Im}[\Phi(\mathbf k)],\nonumber
\end{eqnarray}
and, for even number multipoles $l=2n$,
\begin{eqnarray}
\lefteqn{a_{lm}=}\label{alm3}\\
&&\frac{(-\imath)^l}{\pi^2} \int\limits^{\infty}_0 dk  \int\limits^{\pi}_0 d\theta_{\mathbf k} \sin\theta_{\mathbf k}\int\limits^{\pi}_0 d\phi_{\mathbf k}\,g_{l}(k) Y^*_{lm}(\hat {\mathbf k})\,\mathrm{Re}[\Phi(\mathbf k)]\nonumber.
\end{eqnarray}
It should be noted that the above equations are simple reformulation of Eq. \ref{alm}, and exactly equal to them.

From Eq. \ref{alm2} and \ref{alm3}, we may see that the odd-parity preference might be produced, provided
\begin{eqnarray}
|\mathrm{Re} [\Phi(\mathbf k)]|\ll|\mathrm{Im} [\Phi(\mathbf k)]|\;\;\;(k\lesssim 22/\eta_0),\label{primordial_odd} 
\end{eqnarray}
where $\eta_0$ is the present conformal time.
Taking into account the reality condition $\Phi(-\mathbf k)= \Phi^*(\mathbf k)$, we may show primordial perturbation in real space is given by:
\begin{eqnarray}
\Phi(\mathbf x)&=&2\int\limits^{\infty}_0 dk  \int\limits^{\pi}_0 d\theta_{\mathbf k} \sin\theta_{\mathbf k}\int\limits^{\pi}_0 d\phi_{\mathbf k}\label{Phi_real}\\
&\times&\left(\mathrm{Re}[\Phi(\mathbf k)]\cos(\mathbf k\cdot \mathbf x)-\mathrm{Im}[\Phi(\mathbf k)]\sin(\mathbf k\cdot \mathbf x)\right). \nonumber
\end{eqnarray}
Noting Eq. \ref{primordial_odd} and \ref{Phi_real}, we find our primordial Universe may possess odd-parity preference on large scales ($2/\eta_0 \lesssim k\lesssim 22/\eta_0$).
The odd-parity preference of our primordial Universe violates large-scale translational invariance in all directions.
However, it is not in direct conflict with the current data on observable Universe (i.e. WMAP CMB data), though it may seem intriguing. 
Considering Eq. \ref{primordial_odd} and \ref{Phi_real}, we find this effect will be manifested on the scales larger than $2\pi\,\eta_0/22\approx 4\,\mathrm{Gpc}$.
However, it will be difficult to observe such large-scale effects in non-CMB observations.
If the odd-parity preference is indeed cosmological, it indicates we are at a special place in the Universe, which may sound bizarre.
However, it should be noted that the invalidity of the Copernican Principle such as our living near the center of void had been previously proposed in different context \citep{Void_DE,Void_SN}.

Depending on the type of cosmological origins (e.g. topology, features in primordial power spectrum and Eq. \ref{primordial_odd}), distinct anomalies are predicted in polarization power spectrum. Therefore, polarization maps of large-sky coverage (i.e. low multipoles) will allow us to remove degeneracy and
figure a cosmological origin, if the parity asymmetry is indeed cosmological.

\section{Discussion}
\label{discussion}
We have investigated the parity asymmetry  of our early Universe, using the newly released WMAP 7 year power spectrum data.
Our investigation shows anomalous odd-parity preference of the WMAP7 data ($2\le l\le22$) at 3-in-1000 level. 
When we account for our posteriori choice on $l_\mathrm{max}$, the statistical significance decreases to 2-in-100 level, but remains significant.

There exist several known CMB anomalies at low multipole, including the parity asymmetry discussed in this paper \citep{Tegmark:Alignment,Multipole_Vector1,Multipole_Vector2,Multipole_Vector3,Multipole_Vector4,Axis_Evil,Axis_Evil2,Axis_Evil3,odd,lowl,lowl_WMAP13} (see \citep{lowl_anomalies} for review). We find it likely there exist a common underlying origin, whether cosmological or not.

We have investigated non-cosmological origins, and ruled out various non-cosmological origins such as asymmetric beams, noise and cut-sky effect.
We have also investigated the WMAP team's simulation, which includes all known systematic effects, and do not find definite association with known systematics.
Among cosmological origins, topological models or primordial power spectrum of feature might provide theoretical explanation, though currently available models do not.
We also find primordial origin requires $|\mathrm{Re} [\Phi(\mathbf k)]|\ll|\mathrm{Im} [\Phi(\mathbf k)]|$ for $k\lesssim 22/\eta_0$, if we consider a simple phenomenologically fitting model. In other words, it requires violation of translation invariance in primordial Universe on the scales larger than $4\,\mathrm{Gpc}$.

Depending on the type of cosmological origins, distinct anomalies are predicted in polarization power spectrum.
Therefore, we will be able to remove degeneracy in cosmological origins, when polarization data of large sky coverage are available.
However, at this moment, it is not even clear, whether the anomaly is due to unaccounted contamination or indeed cosmological.
Nonetheless, we may be able to resolve the mystery of the large-scale odd-parity preference, when data from the Planck surveyor are available. 

\section{Acknowledgments}
We are grateful to the anonymous referee for thorough reading and helpful comments, which leads to significant improvement of this work.
We thank Eiichiro Komatsu, Paolo Natoli and Dominik Schwarz for useful discussion.
We acknowledge the use of the Legacy Archive for Microwave Background Data Analysis (LAMBDA). 
We acknowledge the use of the simulated CMB maps of asymmetric beams produced by Wehus et al. \citep{asymmetric_beam}.
Our data analysis made the use of HEALPix \citep{HEALPix:Primer,HEALPix:framework}.  
This work is supported in part by Danmarks Grundforskningsfond, which allowed the establishment of the Danish Discovery Center.
This work is supported by FNU grant 272-06-0417, 272-07-0528 and 21-04-0355.

\begin{appendix}
\section{Statistical Properties of CMB anisotropy}
\label{CMB}
The CMB temperature anisotropy over a whole-sky is conveniently decomposed in terms of spherical harmonics $Y_{lm}(\theta,\phi)$ as follows:
\begin{eqnarray}
T(\hat{\mathbf n})=\sum_{lm} a_{lm}\,Y_{lm}(\hat{\mathbf n}),
\end{eqnarray}
where $a_{lm}$ is a decomposition coefficient, and $\hat{\mathbf n}$ is a sky direction.
Decomposition coefficients are related to primordial perturbation as follows:
\begin{eqnarray}
a_{lm}&=&4\pi (-\imath)^l \int \frac{d^3\mathbf k}{(2\pi)^3} \Phi(\mathbf k)\,g_{l}(k)\,Y^*_{lm}(\hat {\mathbf k}),\label{alm}
\end{eqnarray} 
where $\Phi(\mathbf k)$ is primordial perturbation in Fourier space, and $g_{l}(k)$ is a radiation transfer function.
For a Gaussian model for primordial perturbation, decomposition coefficients satisfy the following statistical properties:
\begin{eqnarray} 
\langle a_{lm} \rangle &=& 0, \\
\langle a^*_{lm} a_{l'm'} \rangle &=& C_l\,\delta_{ll'}\delta_{mm'},
\end{eqnarray}
where $\langle\ldots\rangle$ denotes the average over the ensemble of universes.
Given a standard cosmological model, Sach-Wolf plateau is expected at low multipoles  \citep{Modern_Cosmology}:
$l(l+1) C_l\sim \mathrm{const}$.
\end{appendix}

\bibliographystyle{unsrt}
\bibliography{/home/tac/jkim/Documents/bibliography}
\end{document}